\documentclass[journal,transmag]{IEEEtran}
\ifCLASSINFOpdf
\else
\fi
\usepackage{mathtools}

\usepackage [english]{babel}
\usepackage [autostyle, english = american]{csquotes}
\MakeOuterQuote{"}

\begin{document}
\title{On-chip learning for domain wall synapse based \\ Fully Connected Neural Network}


\author{\IEEEauthorblockN{Apoorv Dankar\IEEEauthorrefmark{1,*},
Anand Verma\IEEEauthorrefmark{1,*},
Utkarsh Saxena\IEEEauthorrefmark{1,*},
Divya Kaushik\IEEEauthorrefmark{1},
Shouri Chatterjee\IEEEauthorrefmark{1}, and
Debanjan Bhowmik\IEEEauthorrefmark{1}}
\IEEEauthorblockA{\IEEEauthorrefmark{1}Department of Electrical Engineering,
Indian Institute of Technology Delhi, New Delhi 110016, India}
% <-this % stops an unwanted space
\thanks{* These authors contributed equally to the work.} 
\thanks{Corresponding author: D. Bhowmik (email: debanjan@ee.iitd.ac.in). Manuscript submitted on Nov 5, 2018 for review}
}

\markboth{}%
{Shell \MakeLowercase{\textit{et al.}}: Bare Demo of IEEEtran.cls for IEEE Transactions on Magnetics Journals}



% for Transactions on Magnetics papers, we must declare the abstract and
% index terms PRIOR to the title within the \IEEEtitleabstractindextext
% IEEEtran command as these need to go into the title area created by
\IEEEtitleabstractindextext{%
\begin{abstract}
Spintronic devices are considered as promising candidates in implementing neuromorphic systems or hardware neural networks, which are expected to perform better than other existing computing systems for certain data classification and regression tasks. In this paper, we have designed a feedforward Fully Connected Neural Network (FCNN) with no hidden layer using spin orbit torque driven domain wall devices as synapses and transistor based analog circuits as neurons. A feedback circuit is also designed using transistors, which at every iteration computes the change in weights of the synapses needed to train the network using Stochastic Gradient Descent (SGD) method. Subsequently it sends write current pulses to the domain wall based synaptic devices which move the domain walls and updates the weights of the synapses. Through a combination of micromagnetic simulations, analog circuit simulations and numerically solving FCNN training equations, we demonstrate "on-chip" training of the designed FCNN on the MNIST database of handwritten digits in this paper. We report the training and test accuracies, energy consumed in the synaptic devices for the training  and possible issues with hardware implementation of FCNN that can limit its test accuracy.
\end{abstract}

}

\maketitle

\IEEEdisplaynontitleabstractindextext

% \fi
\IEEEpeerreviewmaketitle

\section{Introduction}
\IEEEPARstart{A}{rtificial} Neural Network (ANN) algorithms are currently being widely used by the machine learning and data sciences community to solve several kinds of data classification and regression problems \cite{LeCun}. These ANN algorithms, inspired by the working of the human brain inherently have memory and computing intertwined in them just like the brain. For example, in a feedforward Fully Connected Neural Network (FCNN) with no hidden layer, signals at the nodes of the input layer are multiplied by specific values called weights and then added up, followed by operation of an activation function on them, which leads to signals at the nodes of the output layer (Fig. 1(a)). Storing these weights in the network constitutes the memory functionality of the algorithm while the calculation of the product between the input signals and the weights, the summation of the products and operation of the activation function on the sum constitute the forward computation functionality of the algorithm. The network is trained to perform specific regression and classification tasks, under the supervised learning scheme, by updating these weights after every iteration on the training examples using the Stochastic Gradient Descent (SGD) method until the error at the output is minimized \cite{LeCun}. When such ANN algorithms are implemented in software, as has been the case currently, the algorithms are still executed on traditional computer hardware in which memory and computing are separate, following the von Neumann architecture \cite{NeuralNetSurvey_Misra, NeuralNetSurvey_Schuman}. Thus the property of memory-computing entanglement inherent in these algorithms cannot be properly utilized.

However in neuromorphic computing systems, specialized hardware, where memory and computing are embedded together, is designed to implement such ANN algorithms. This can enhance the performance of the computing system with respect to speed and energy consumption \cite{NeuralNetSurvey_Misra, NeuralNetSurvey_Schuman,NeuralNetComparison_Diamond}. Fig. 1(b) shows analog hardware implementation of single layer FCNN. Weights are stored and updated as conductances of the synaptic devices- devices that mimic synapses in the brain \cite{Kaushik_BioMedCircuit}. Forward computation takes place by adding currents flowing out of these synaptic devices, arranged in a crossbar architecture, followed by operation of a tan-sigmoid activation function on them through a transistor based circuit, partly mimicking the neurons in the brain. 

Spintronic devices, owing to their non-volatile nature, are particularly suitable as synaptic devices in neuromorphic computing systems \cite{Kaushik_BioMedCircuit,Kaushik_AppPhysRev,Grollier}. If a domain wall can be created in the ferromagnetic metal layer (free layer) of a heavy metal/ ferromagnetic metal (free layer)/ oxide / ferromagnetic metal (fixed layer) heterostructure corresponding to a Magnetic Tunnel Junction (MTJ) device, as long as the domain wall does not move, Tunneling Magneto Resistance (TMR) of the MTJ structure does not change (Fig. 2(a)) \cite{Ikeda,Emori,Ryu,Bhowmik,skyrmionMTJ1}. Hence the device can act as synapse in hardware ANN and store the corresponding weight of the synapse as its conductance. Further, in order to update the weight after every iteration with the goal of training the network, "write" current pulses can be applied through the heavy metal layer of the device such that spin orbit torque from the current pulses can move the domain wall and bring about the required change in conductance of the device and hence the weight \cite{Ikeda,Emori,Ryu,Sampaio,Fert}. However a dedicated circuitry making use of the SGD method \cite{LeCun} will be needed to generate those suitable current pulses that can eventually train the network. This scheme of training the network in hardware along with the forward computation is known as "on-chip learning". 

\begin{figure}
\centerline{\includegraphics[width=3in]{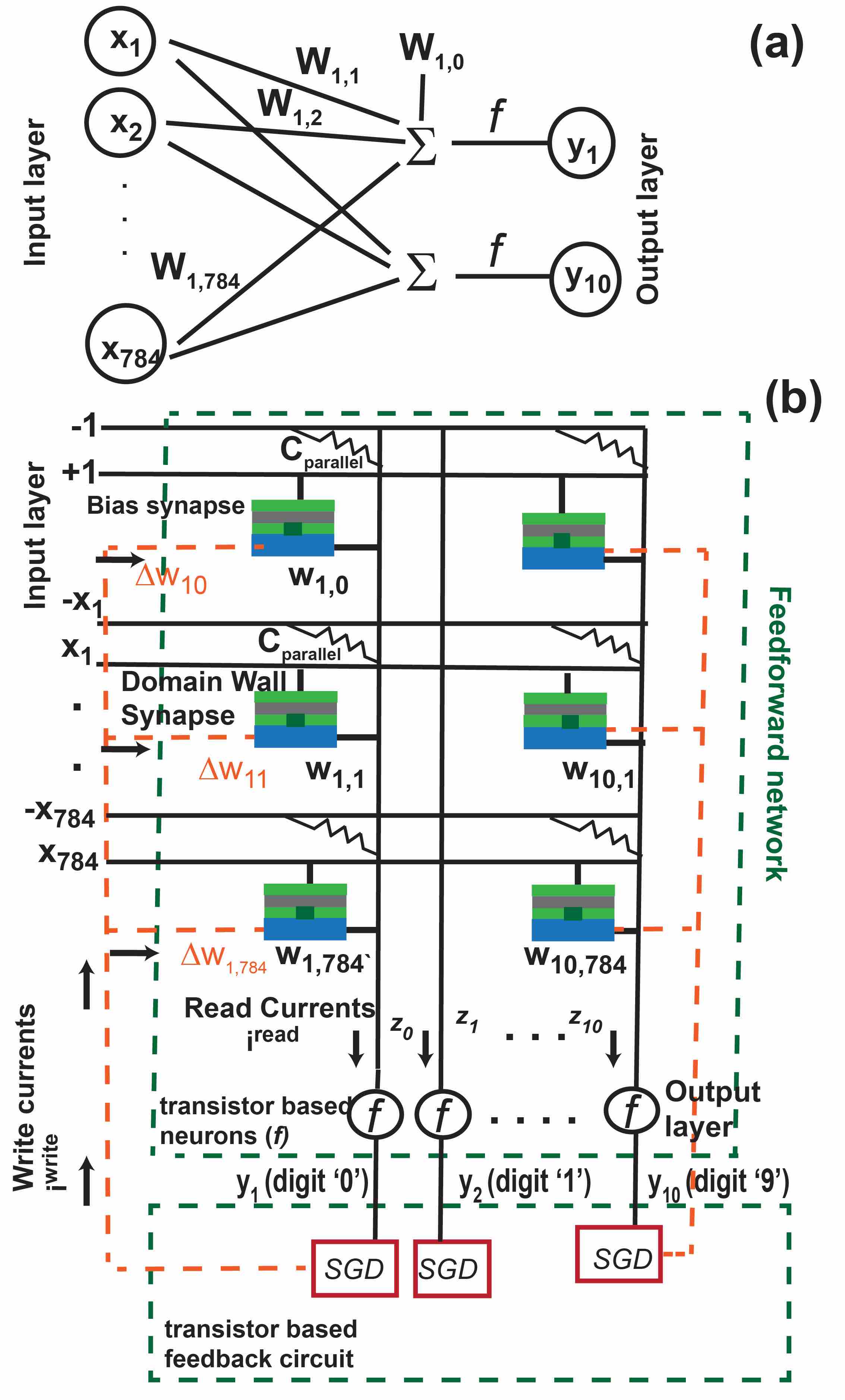}}
\caption{(a) Schematic of Fully Connected Neural Network (FCNN) without a hidden layer. Each $x$ corresponding to input feature (intensity of each of the 28 x 28 =784 pixels of images from MNIST character dataset say), $w$ corresponds to weight matrix, $f$ activation function and each $y$ corresponds to output of the network. For the node corresponding to the digit input image belongs to (0-9) $y$ = 1, else -1.  (b) Implementation of FCNN in analog hardware. Domain wall based synaptic device stores weight $w$, transistor based neuron circuit evaluates $f$ and transistor based feedback circuit updates $w$ using Stochastic Gradient Descent (SGD) method.}
\end{figure}

To the best of our knowledge, simulation of "on-chip learning" on a spintronic FCNN system has not been reported before. Though some simulation based reports of ferromagnetic domain wall device based implementation of FCNN have been published recently, they do not implement "on-chip learning" in their networks, i.e. the weight update method is not implemented in hardware \cite{Kaushik_BioMedCircuit,Kaushik_BioCAS} . Essentially, in those reports, several iterations of forward computation and weight update are first run on a standard computer to obtain the final weight values of the synapses for the trained network. Then current pulses are applied on the domain wall based synaptic devices such that their conductances are proportional to the final synaptic weights. Subsequently the forward computation is implemented in hardware. Thus learning is "off-chip" in this hardware implementation of neural network and hence proper advantage is not taken from the memory- computing intertwining present in hardware ANN. However in this paper, we employ a combination of micromagnetic simulations and transistor based circuit simulations to implement "on-chip learning" in such spintronic neural networks. We use spin orbit torque driven ferromagnetic domain wall devices in \cite{Kaushik_BioMedCircuit, Saxena} as synaptic devices in our networks. We use Metal Oxide Semiconductor Field Effect Transistor (MOSFET) and such MOSFET based operational amplifier circuitry to implement the activation function (neuron) and generate current pulses in feedback, using the SGD method,  that move the domain wall in the synaptic devices and change their weights after every iteration (Fig. 1(b)). It is to be noted that spintronic devices, domain wall based devices in this case, are only used as synapse/ memory elements in our network owing to their non-volatility.  Implementing a synapse with existing transistor technology is problematic because of the large number of transistors needed to represent one synapse (around 6-8) and the high power consumption to retain a weight value at the synapse during training of the network because a transistor is a volatile device \cite{Kaushik_IEEEReview}. However for every other functionality in the network where non-volatility is not needed, be it the neuron or SGD calculation circuitry, transistor based circuits are used in our work since existing technology facilitates much easier fabrication of silicon based transistor circuitry compared to magnetic materials based spintronics circuitry \cite{SpintronicsRoadmap}. 

It is also to be noted that though "on-chip learning" for domain wall synapse based ANN is simulated in \cite{Kaushik_AppPhysRev,Kaushik_STDP1} the ANN simulated there is of spiking type unlike ours. Also unlike our work, the synapse there follows a local learning rule- Spike Time Dependent Plasticity (STDP) for weight update   \cite{BiPoo,Skyrmion_WangKang1}, the neuron follows Leaky Intergrate Fire (LIF) model \cite{Skyrmion_WangKang2} and an unsupervised learning is followed for training \cite{DiehlSpiking}. Though such STDP enabled spiking network is closer to the functioning of the brain, the machine learning and data sciences community currently use non-spiking ANN with SGD method based weight update much more than STDP enabled spiking ANN for various tasks. Hence it is important to study "on-chip learning" of such FCNN in spintronic hardware which we have done in this paper, 

Section II discusses the micromagnetic simulations performed to obtain the current controlled conductance characteristics of the domain wall based devices, used as synapses in the implemented FCNN. Section III discusses the design of the FCNN in hardware and how the forward computation and backpropagation algorithm are executed in the it. Section IV evaluates the performance of the network when it is trained and tested on the MNIST database of handwritten digits, abundantly used by the machine learning community to benchmark the performance of different algorithms. We see from the signal flow in the circuit for different inputs that the operational amplifier based SGD computation circuitry designed here is indeed capable of sending the appropriate current pulses to the spintronic synaptic devices and update their weights to successfully train the network. In Section V we summarize and comment on our results and conclude the paper.

\section{Synaptic Device Characteristic}

Spin orbit torque driven ferromagnetic domain wall device was proposed as synaptic element in hardware ANN by Sengupta \textit {et al.} in \cite{Kaushik_BioMedCircuit}. In this work, we simulate such a device on micromagnetic simulation package "mumax3" \cite{mumax} and obtain its synaptic characteristic- conductance of the vertical Magnetic Tunnel Junction (MTJ) structure as a function of "write" current flowing horizontally through the heavy metal layer (Fig. 2(a)) that can move the domain wall through the application of spin orbit torque on the magnetic moments inside the wall \cite{Saxena}. 

\begin{figure}
\centerline{\includegraphics[width=3in]{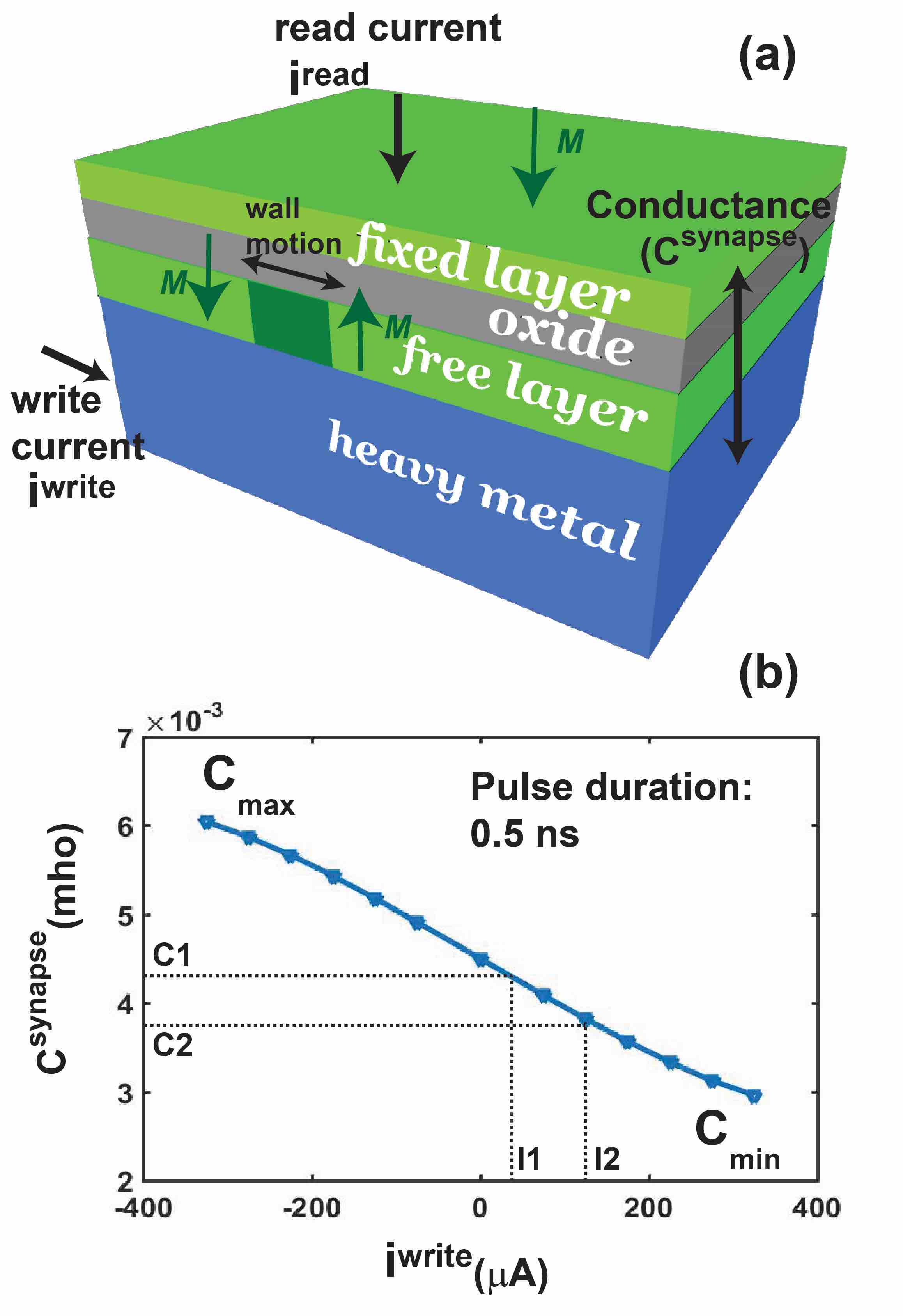}}
\caption{(a)Schematic of domain wall based synaptic device (b) Conductance of the vertical MTJ structure in the device ($C^{synapse}$) after application of different magnitudes of write current ($i_{write}$) pulse horizontally through the heavy metal layer is computed through micromagnetic simulations and then plotted.}
\end{figure}

The lateral device dimensions of our synaptic device (Fig. 2(a)) are taken to be 500 nm in length and 50 nm in width. The ferromagnetic free layer, in which the domain wall is formed, is taken to be 1 nm thick in our micromagnetic simulations. We take saturation magnetization ($M_s$) = $8 \times 10^{5}$ A/m , exchange correlation constant  (A) = $3 \times 10^{-11}$ J/m  and damping factor = 0.02 throughout the free layer. Perpendicular Magnetic Anisotropy (PMA) constant (K) is taken to be $8 \times 10^{5} J/m^3$ considering perpendicularly magnetized CoFeB/MgO structure \cite{Bhowmik}. Dzyalonshinskii Moriya Interaction (DMI) is taken to be $3 \times 10^{-3}$ J/m$^2$ \cite{Sampaio}. Neel domain wall is stabilized at such value of DMI in our simulations. 

Dynamics of the domain wall formed in the ferromagnetic free layer is simulated in the presence of vertical spin current, that acts upon the magnetic moments due to charge current flowing horizontally in the heavy metal layer under the ferromagnetic free layer (Fig. 2(a)). Spin current density= spin Hall angle $\times$ charge current density, where charge current density= charge current / Cross-sectional area, and cross-sectional area = width (50 nm) $\times$ thickness of heavy metal layer \cite{Bhowmik, Liu_Science}. To obtain this expression for spin current density, it is assumed that the thickness of the heavy metal layer is greater than the spin diffusion length inside the heavy metal so that the spin current becomes independent of the thickness of the heavy metal layer \cite{Liu_Science, Zhang}. We consider platinum (Pt) as the heavy metal here. Since spin diffusion length in Pt has been reported experimentally to be 2-4 nm \cite{SpinDiffLength_1,SpinDiffLength_2} and the thickness of the heavy metal layer considered here is 10 nm, the assumption holds true in this case. The value of spin Hall angle of Pt is considered to be 0.07 in our work \cite{Liu_PRL1,Liu_PRL2}. 

Conductance vs "write" current characteristics of the simulated device is shown in Fig. 2(b). Starting from domain wall at the center of the device, current pulse of 0.5 ns in duration and about 400 $\mu A$ in magnitude is needed to move the domain wall all the way to one edge, corresponding to the anti-parallel alignment of magnetic moments of the free and fixed layer and hence minimum conductance of the MTJ ($C_{min}$). About - 400 $\mu A$ current is needed to move the domain wall to the other edge, corresponding to the parallel alignment of magnetic moments of the free and fixed layer and hence maximum conductance ($C_{max}$) (Fig. 2(b)). Intermediate conductance values are obtained by applying current pulses of magnitude between -400 $\mu A$ and + 400 $\mu A$ and duration 0.5 ns. Such conductance values correspond to the different values of weight that the device can store as a synaptic element in the neural network.  For conductance calculation, Resistance- Area (RA) product of the MTJ is taken to be $4.04 \times 10^{-12}$ $\Omega m^2$ \cite{RAproduct} and the TMR value is taken to be 120 $\%$ \cite{Ikeda}. This leads to values of $C_{min}$ and $C_{max}$ (Fig. 2(b)). Intermediate conductance values are calculated using the expression: \emph{ $C^{synapse}$ = ($C_{min}$ + $C_{max}$ )/2 + ($C_{min}$ - $C_{max}$ )*$m_{avg}$/2}, where $m_{avg}$ represents the average perpendicular component of the magnetization of the free layer.

It is to be noted that the device is non-volatile. Once current $I1$ is applied to obtain conductance $C1$, corresponding to weight $w1$, and then removed, conductance remains to be $C1$. Thus weight of the corresponding synapse in the network continues to be $w1$. However in order to train the network the weight may need to be updated to $w2$ at a certain iteration. In that case, a current pulse of strength $I2-I1$ needs to be applied horizontally through the heavy metal layer of the device to change the conductance of the MTJ structure to $C2$, corresponding to weight $w2$. Thus, to change conductance by $\Delta C^{synapse}=C2-C1$, a small current $i^{write}=I2-I1$ needs to be applied.  If the device isn't non-volatile a much larger current $I2$ will be needed for the same weight update from $w1$ to $w2$.

Thus for our domain wall based synaptic device simulated here, 

\begin{equation}
i^{write}=\frac{\partial i^{write}}{\partial C^{synapse}}\Delta C^{synapse}
\end{equation}

From the micromagnetic simulations we perform on "mumax3", $\frac{\partial i^{write}}{\partial C^{synapse}}=-2.1\times10^{5}\mu$A-$\Omega$ when duration of the current pulse = 0.5 ns (Fig.2(b)). When duration of the pulse is the 5 ns, smaller magnitudes of current pulses are needed to achieve the same conductance states because domain wall velocity is proportional to current density in our simulations, which is also confirmed experimentally \cite{Emori,Ryu,Saxena,Martinez}. From our micromagnetic simulations, 
$\frac{\partial i^{write}}{\partial C^{synapse}}$ hence turns out to be $6\times10^{3}\mu$A-$\Omega$ when duration of the pulse = 5 ns.

In the following section, we discuss how several such synaptic devices can form a feedforward FCNN to generate "read" currents at the nodes of the output layer of the network and how a transistor based feedback circuitry we designed applies "write" currents on the heavy metal layers of the synaptic devices to change their weights by required amounts at every iteration to eventually train the network.

\section{Design of feedforward network and feedback circuitry}

\subsection{Feedforward network}
We simulate a cross-bar architecture of spin orbit torque driven domain wall synaptic devices of Fig. 2(a) to form a feedforward Fully Connected Neural Network (FCNN) as shown in Fig. 1(b) \cite{Kaushik_BioMedCircuit, Saxena}. Architecture of a standard FCNN without any hidden layer is shown in Fig. 1(a). In this paper, we train the FCNN to identify digits from 0 to 9 from the standard MNIST handwritten digit database \cite{MNIST}. So number of nodes in input layer = number of pixels of each input image of a digit = 28 x 28 =784. Input to the nodes of the input layer ${\{x_1,x_2,x_3.....x_{784}\}}$ correspond to the intensities of the pixels. Number of nodes in output layer = number of digits = 10. The desired output when input image is of digit 0 is given by ${\{Y_1,Y_2,Y_3,....Y_{10}\} = \{1,-1,-1,....-1\}}$, for digit 1 is given by ${\{Y_1,Y_2,Y_3,....Y_{10}\} = \{-1,1,-1,....-1\}}$ and so on. The target of training the network is  such that for a given input ${\{x_1,x_2,x_3.....x_{784}\}}$ the output 
${\{y_1,y_2,y_3,....y_{10}\}}$ at the output layer of the network matches the desired output ${\{Y_1,Y_2,Y_3,....Y_{10}\}}$. Once the network is trained and gives a high accuracy for an input from the training set (training accuracy) its accuracy needs to be tested on a fresh set of inputs (test accuracy).
Following the standard FCNN training algorithm \cite{LeCun,Haykin},
output at any node n is given by: 
\begin{multline}
y_n=f(z_n) = \frac{2}{1+e^{-\lambda z_n}}-1;\\
z_n = w_{n,1}x_1 + w_{n,2}x_2 + ...... w_{n,784}x_{784}+w_{n,0}\\
=(\Sigma_{m=1}^{m=784}w_{n,m}x_m)+ w_{n,0}
\end{multline}
where $f$  is the activation function, $w_{n,1}, w_{n,2}..... w_{n,784},w_{n,0}$-s are the synaptic weights, $w_{n,0}$ being the bias weight. Equation (2) is essentially a matrix-vector multiplication, matrix being weight $w$ and vector being $x$, to obtain vector $z$  followed by operation of a non linear function (f)- "tan sigmoid" in this case \cite{Haykin}- on every element of z to obtain vector y. 

In order to implement equation (2) in hardware, voltages $\{V^{in}_1,V^{in}_2,V^{in}_{3},........V^{in}_{784}\} = V_{source}*\{x_1,x_2,x_3,.....x_{784}\}$ are applied on the cross-bar architecture of the domain wall based synaptic devices as shown in Fig. 1(b).  Since conductance of the devices ($C^{synapse}$) only takes positive values and ranges between $C_{min}$ and $C_{max}$ (Fig. 2(b)) while the corresponding weights $w$ can take both positive and negative values, an extra conductance ($C_{parallel}$) is added in parallel to each of the synapses and negative of the voltage applied on the synapse is applied on it. The relation between conductance of a synapse connecting input node $m$ with output node $n$ and its corresponding weight $w_{n,m}$ is as follows: 

\begin{equation}
C^{synpase}_{n,m}=(\frac{C_{max}-C_{min}}{2w_{max}})w_{n,m} + C_{parallel}
\end{equation}

where $w_{max}$ is the magnitude of the maximum weight value in the network. 
\begin{equation}
C_{parallel}=\frac{C_{max}+C_{min}}{2}
\end{equation}

For a voltage $V_{source}x_m$ applied on the input node $m$, current flowing through the combination of domain wall synapse device and extra conductance ($C_{parallel}$), connecting input node $m$ with output node $n$,  is given by : 
\begin{multline}
i^{read}_{n,m}=(C^{synapse}_{n,m}V_{source}x_m)-(C_{parallel}V_{source}x_m) \\
= \frac{C_{max}-C_{min}}{2w_{max}}w_{n,m}V_{source}x_{m}
\end{multline}

for m = 1 to 784. It is to be noted that this current, which we call "read" current in this paper, flows through the vertical MTJ structure of the synaptic device (Fig. 2(a)) and is hence proportional to the conductance of the MTJ. It is not the "write" current that flows horizontally through the heavy metal layer to move the domain wall and change the conductance of the MTJ (Fig. 2(a)). Magnitude of "read" current is proportional to $V_{source}$. Value of $V_{source}$ is chosen such that the maximum value of "read" current flowing through the synaptic device is not large enough to move the domain wall and change the weight value it is storing. For the circuit we design here, 
$V_{source}$ is chosen to be 1 mV.

Corresponding to the bias weight $w_{n,0}$ there is a bias synapse with conductance $C^{synapse}_{n,0}$ (Fig. 1(b)) and "read" current flowing through it is given by: 

\begin{multline}
i^{read}_{n,0}=(C^{synapse}_{n,0}V_{source})-(C_{parallel}V_{source}) \\
= \frac{C_{max}-C_{min}}{2w_{max}}w_{n,0}V_{source}
\end{multline}

At the output node $n$, "read" currents from all connected synapses add up following Kirchoff's Current Law (Fig. 1(b)) to yield the total "read" current:

\begin{multline}
I^{read}_{n}=\Sigma_{m=0}^{m=784}i_{n,m}=\\ \Sigma_{m=1}^{m=784}\frac{C_{max}-C_{min}}{2w_{max}}w_{n,m}V_{source}x_{m}\\
+\frac{C_{max}-C_{min}}{2w_{max}}w_{n,0}V_{source}\\
=(V_{source}\frac{C_{max}-C_{min}}{2w_{max}})z_n
\end{multline}

Thus the matrix-vector multiplication of equation (2) is accomplished in hardware as shown by equation (7), with an extra scaling factor $V_{source}\frac{C_{max}-C_{min}}{2w_{max}}$ coming from the circuit implementation.

\begin{figure}
\centerline{\includegraphics[width=3in]{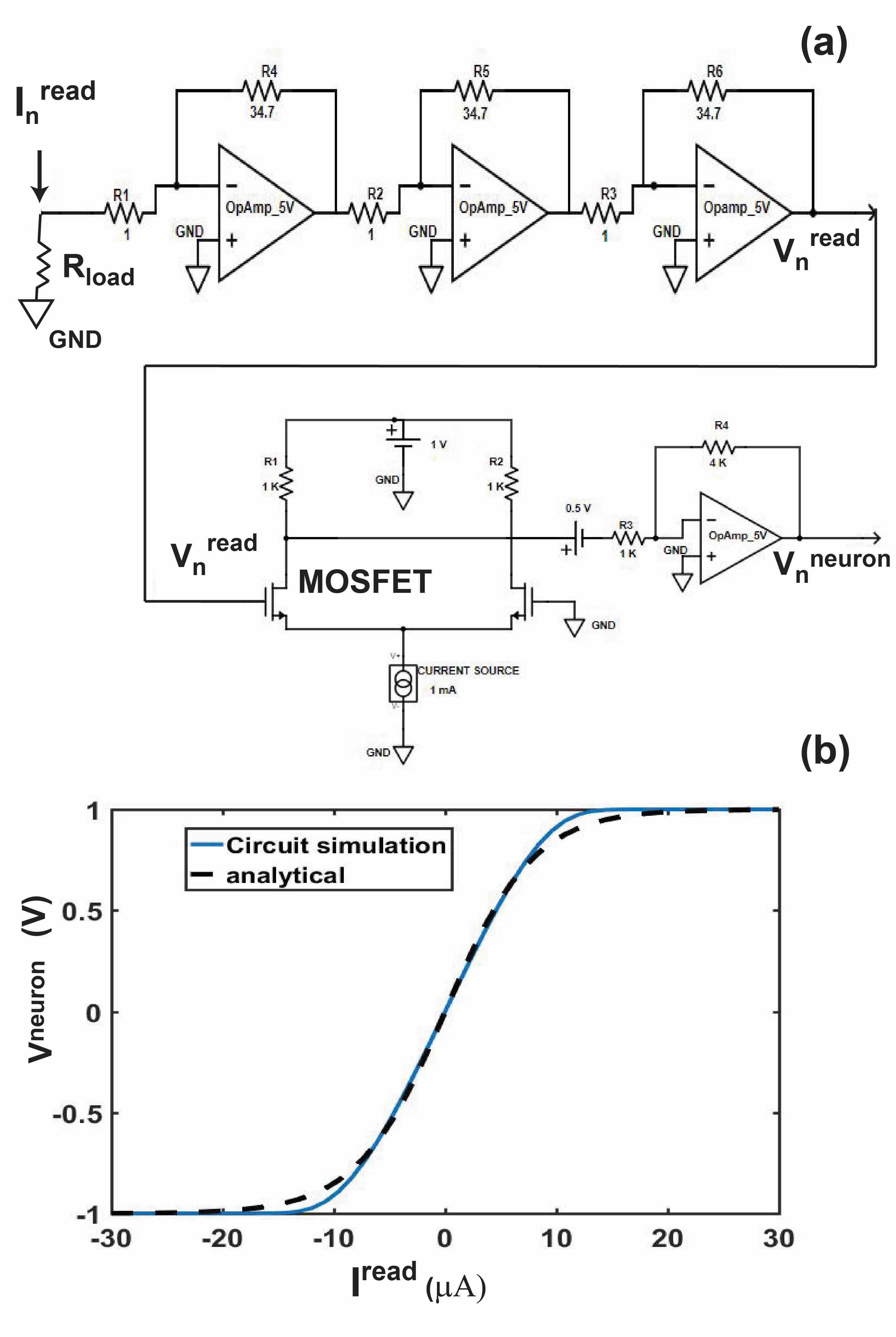}}
\caption{(a) Transistor based neuron circuit present at each output node of the feedforward network, consisting of an op-amp amplifier and differential circuit that takes voltage as input. (b) Output voltage of neuron circuit ($V^{neuron}$) as a function of input current ($I^{read}$), obtained from circuit simulation as well as analytical calculation (equation 9), is plotted.}
\end{figure}

\begin{figure}
\centerline{\includegraphics[width=3in]{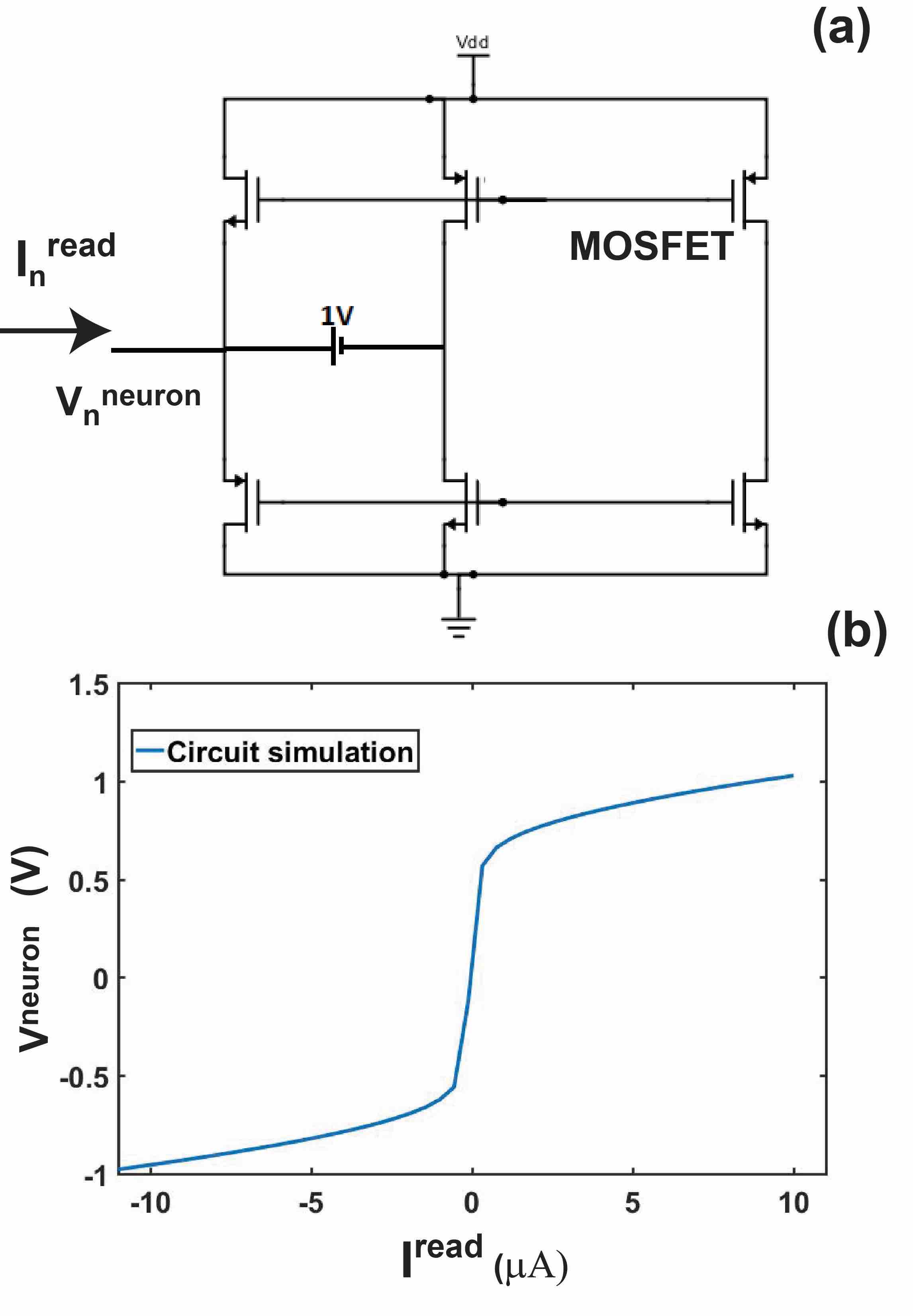}}
\caption{(a) Alternative transistor based neuron circuit consisting of a differential circuit that takes current as input. (b) Output voltage of neuron circuit ($V^{neuron}$) as a function of input current ($I^{read}$), obtained from circuit simulation is plotted.}
\end{figure}

The activation function $f$ of equation (1) is implemented at each output node of the circuit (Fig.1(b)) using a transistor based "neuron" circuit of Fig. 3(a).
The net "read" current at each output node $n$, given by equation (7), is first passed through a very low resistance- $R_{load}$ (1 $\Omega$ in this case). $R_{load}$ is chosen so low so that the voltage at the output node stays close to 0 and the  expression for "read" current in equation (7) remains valid. The voltage across $R_{load}$ is next amplified through an op-amp (transistor based high gain amplifier) circuit \cite{circuitbook} to eliminate the extra scaling factor in equation (7), to generate an output voltage ($V^{read}_{n}$): 

\begin{multline}
V^{read}_{n}= \\
(\frac{1}{V_{source}\frac{C_{max}-C_{min}}{2w_{max}}R_{load}\lambda_{circuit}})I^{read}_{n}R_{load}
\end{multline}

This voltage $V^{read}_{n}$ is next fed to one of the two inputs of the MOSFET based differential amplifier circuit of Fig. 3(a), designed by us, which operates the tan-sigmoid function on it \cite{circuitbook,NeuronCircuit}. The factor $\lambda_{circuit}$ arises in equation (8) since the $\lambda$ parameter in our FCNN algorithm of equation (1) is 1 while the same factor for the differential amplifier based tan-sigmoid circuit ($\lambda_{circuit}$) we design is 6. The amplification factor of equation (8) ($\frac{1}{V_{source}\frac{C_{max}-C_{min}}{2w_{max}}R_{load}\lambda_{circuit}}$) turns out to be 42000. Such high amplification needed is carried out with three stages of op-amps as shown in Fig. 3(a).

The output of the differential amplifier circuit, and hence the "neuron" circuit of Fig. 3(a) is expected to be: 

\begin{multline}
V^{neuron}_{n}=\frac{2}{1+e^{-\lambda_{circuit} V^{read}_{n}}}-1=\\
\frac{2}{1+e^{-\lambda_{circuit}(\frac{1}{V_{source}\frac{C_{max}-C_{min}}{2w_{max}}R_{load}\lambda_{circuit}})I^{read}_{n}R_{load}}}-1\\
=\frac{2}{1+e^{-\lambda z_{n}}}-1=y_n
\end{multline}

We plot $V^{neuron}_{n}$ as a function of $I^{read}_{n}$ from equation (9) using the appropriate values of parameters in Fig. 3(b) (analytical plot).  We next simulate the circuit of Fig. 3(a) on Cadence Virtuso circuit simulator to also obtain $V^{neuron}_{n}$ as a function of $I^{read}_{n}$, as plotted in Fig. 3(b). United Microelectronics Corporation (UMC) 65 nm technology node library is used. Length of transistor is chosen to be 80 nm and width 60 nm.  We see that the analytical plot and the plot from circuit simulations match quite well, which means the differential amplifier circuit works as per our expectations in terms of generating the tan-sigmoid function. Also from equation (9) we see that $V^{neuron}_{n}$ in hardware represents output at node n ($y_n$) in the FCNN of equation (9), without any extra factor coming from the hardware.

The circuit of Fig. 3(a) has the drawback that it needs a high voltage gain amplifier circuit which can give erroneous result in a noisy environment. An alternative circuit is presented in Fig. 4(a) \cite{NeuronCircuit2}. Here, the transistor based differential circuit takes the "read" current as input directly unlike the circuit of Fig. 3(a). Hence it is not needed to make the "read" current flow through a load resistance and amplify the voltage across the resistance unlike the previous case. Thus the very high gain op-amp circuit of Fig. 3(a) is avoided here.  Output voltage of the circuit $V^{neuron}_{n}$ as a function of input "read" current $I^{read}_{n}$, obtained from simulation of this circuit on Cadence Virtuoso simulator, is plotted in Fig. 4(b).
\subsection{Feedback Circuitry}

\begin{figure*}
\centering
\centerline{\includegraphics[width=6.0in]{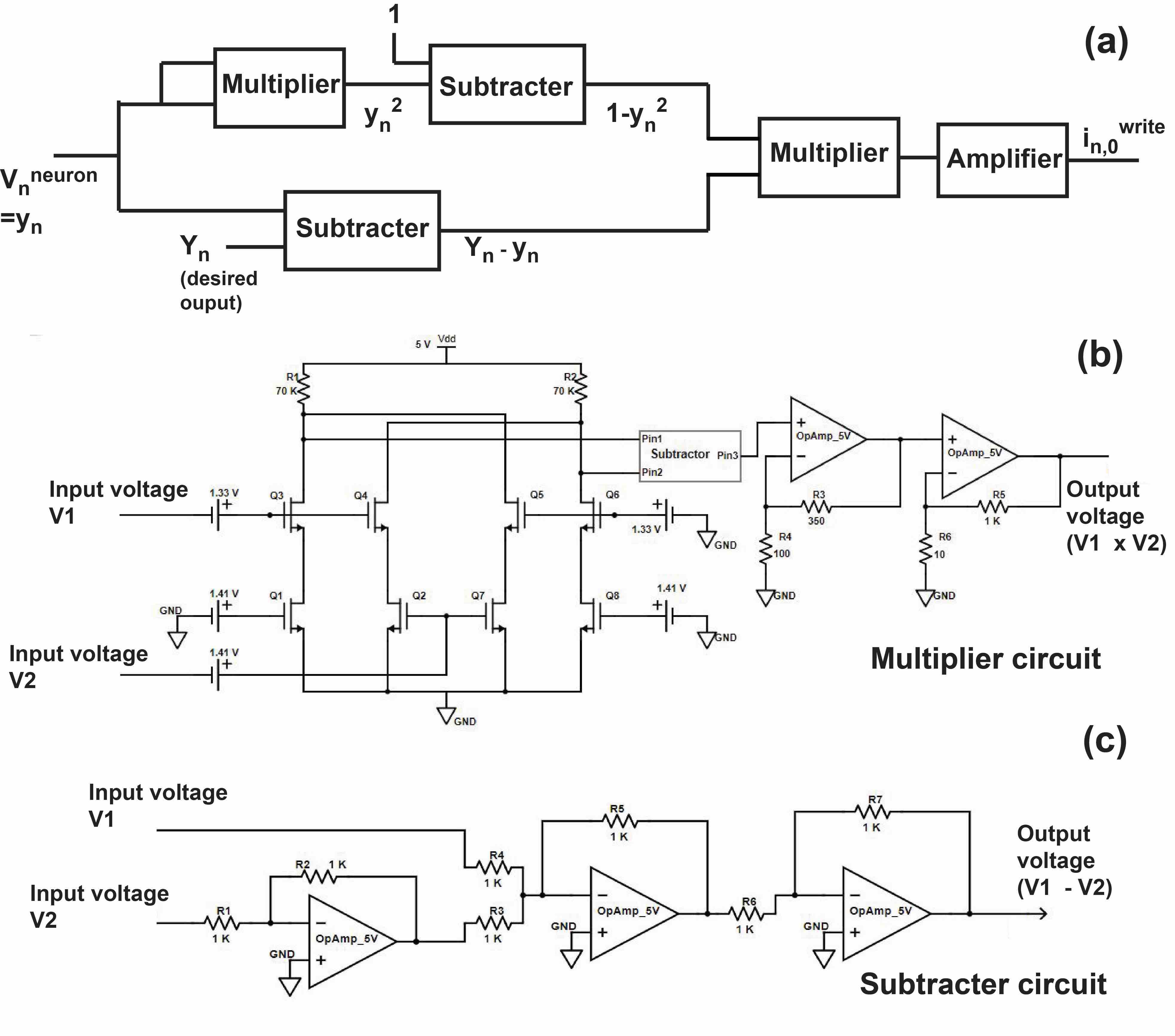}}
\caption{(a) Schematic of the feedback circuit at every node of the FCNN which evaluates the change in synaptic weight  using SGD method (equation 11 and 12) and  generates write current pulse needed to bring about the change in weight is shown here.  It is to be noted that the write current generated here is for the bias synapse connected to that output node. For synapses that connect the input nodes with this output node, the write current generated here has to be multiplied with input signal at the corresponding input node as given in equation (11). (b) Transistor based implementation of multiplier circuit we design here. (c) op-amp based implementation of subtracter circuit we design here.}
\end{figure*}

We next describe how the weights of the ANN are updated to train the network in  the Stochastic Gradient Descent (SGD) algorithm followed here. For a given training example ${\{x_1,x_2,x_3.....x_{784}\}}$, output ${\{y_1,y_2,y_3,....y_{10}\}}$ is generated using the feedforward computation of equation (1). Since a supervised learning algorithm is followed, expected output for that training example ${\{Y_1,Y_2,Y_3,....Y_{10}\}}$ is known and error at output node $n$ is calculated as follows:
\begin{equation}
\epsilon_{n}=\frac{1}{2}{(Y_n-y_n)^{2}}
\end{equation}
Following the SGD method \cite{LeCun,Haykin}, weight of synapse connecting input node $m$ with output node $n$ is updated as follows between iteration $i$ and $i+1$:

\begin{multline}
w^{i+1}_{n,m}=w^{i}_{n,m}-\Delta w_{n,m}\\
=w^{i}_{n,m}-\eta\frac{\partial \epsilon_n}{\partial w_{n,m}}\\
=w^{i}_{n,m}-\eta(Y_n-y_n)(-\frac{\partial y_n}{\partial w_{n,m}}) \\
=w^{i}_{n,m}-\eta(Y_n-y_n)(-\frac{\partial y_n}{\partial z_{n}})(\frac{\partial z_{n}}{\partial w_{n,m}}) \\
=w^{i}_{n,m}-\frac{\eta\lambda}{2}(Y_n-y_n)(1-y^2_n)x_m
\end{multline}

and weight of the bias synapse for output node $n$ is updated as follows:

\begin{multline}
w^{i+1}_{n,0}=w^{i}_{n,0}-\Delta w_{n,0}\\
=w^{i}_{n,0}-\eta\frac{\partial \epsilon_n}{\partial w_{n,0}}\\
=w^{i}_{n,0}-\eta(Y_n-y_n)(-\frac{\partial y_n}{\partial w_{n,0}}) \\
=w^{i}_{n,0}-\eta(Y_n-y_n)(-\frac{\partial y_n}{\partial z_{n}})(\frac{\partial z_{n}}{\partial w_{n,0}}) \\
=w^{i}_{n,0}-\frac{\eta\lambda}{2}(Y_n-y_n)(1-y^2_n)
\end{multline}

where $\eta$ is the learning rate, equal to 0.1 in our simulations.

The training sample is changed at every iteration to exhaust all examples in the training set. Then this process is repeated several times, each repetition being called an epoch. Thus, total number of iterations= number of epochs $\times$ number of training samples.

Corresponding to the calculation of ${\{y_1,y_2,y_3,....y_{10}\}}$ in the algorithm for the training example ${\{x_1,x_2,x_3.....x_{784}\}}$, voltages ${\{V^{neuron}_1,V^{neuron}_2,V^{neuron}_3.....V^{neuron}_{784}\}}$ are generated at the output nodes of the corresponding feedforward computation circuit of Fig. 1(b) as described in the previous subsection. At each output node $n$ $V^{neuron}_{n}=y_{n}$ as we have already shown. Now, this $V^{neuron}_n$ is fed to the feedback circuit at that node (Fig. 1(b)) which evaluates $\Delta w_{n,m}$ at that iteration for all the synapses connecting that output node $n$ with all input nodes from $m=1$ to $m=784$ and $\Delta w_{n,0}$ the bias synapse. Details of the feedback circuit that we have designed for the purpose and simulated on Cadence Virtuoso are shown in Fig. 5(a). $y_n$ is split into two branches. At one branch it is subtracted from the desired  output signal $Y_n$, using the op-amp based subtractor circuit of Fig. 5(c), to generate the $Y_n-y_n$ term. At the other branch it is multiplied with itself using MOSFET based Gilbert cell multiplier circuit of Fig. 5(b) \cite{GilbertMultiplier} and then subtracted from a constant voltage of 1 V using the subtractor circuit of Fig. 5(c) to generate  the $1-y^2_n$ term. Then the two terms are multiplied using the Gilbert cell multiplier of Fig. 5(b) and amplified by the factor of $\frac{\eta\lambda}{2}$ to generate $\Delta w_{n,0}$. Then it is multiplied by the input $V_m$ at each input node $m$ (scaled by $V_{source}$) to generate $\Delta w_{n,m}$, as shown in Fig. 5(a).   

Using equation (3) the required change in the conductance of the corresponding domain wall based synaptic device is given by:
\begin{multline}
\Delta C^{synapse}_{n,m} = (\frac{C_{max}-C_{min}}{2w_{max}})\Delta w_{n,m}; \\
\Delta C^{synapse}_{n,0} = (\frac{C_{max}-C_{min}}{2w_{max}})\Delta w_{n,0} 
\end{multline}

From equation (1) the "write" current that needs to be applied through the heavy metal layer of the synaptic devices to bring about the required change in conductance, and hence change in weight, is given by:
\begin{multline}
i^{write}_{n,m}=\frac{\partial i^{write}}{\partial C^{synapse}}\Delta C^{synapse}_{n,m} = \\ \frac{\partial i^{write}}{\partial C^{synapse}}(\frac{C_{max}-C_{min}}{2w_{max}})\Delta w_{n,m}; \\
i^{write}_{n,0}=\frac{\partial i^{write}}{\partial C^{synapse}}\Delta C^{synapse}_{n,0} = \\ \frac{\partial i^{write}}{\partial C^{synapse}}(\frac{C_{max}-C_{min}}{2w_{max}})\Delta w_{n,0}
\end{multline}

The feedback circuit of Fig. 5(a) amplifies the  $\Delta w_{n,m}$ and $\Delta w_{n,0}$ it computes with the appropriate scaling factor in equation (14) to compute the corresponding "write" currents $i^{write}_{n,m}$ and $i^{write}_{n,0}$. Then it applies voltage $V^{write}_{n,m}=R_{write}i^{write}_{n,m}$ on the heavy metal layer of synaptic device connecting input node $m$ with output node $n$ and $V^{write}_{n,0}=i^{write}_{n,0}R_{write}$ on heavy metal layer of bias synapse device at output node $n$ (Fig. 1(b)) to bring about the changes $\Delta w_{n,m}$ and $\Delta w_{n,0}$ in the weights of the corresponding synapses. $R_{write} $ is the resistance of the heavy metal layer. This process repeated over a certain number of iterations trains the network and "on-chip learning" is achieved.

In the following section, we show the results we obtain connected to our simulation of "on-chip learning" for the designed hardware FCNN on the MNIST character dataset using the method described in this section.

\section{Performance of designed hardware network}

\begin{figure}
\centerline{\includegraphics[width=3in]{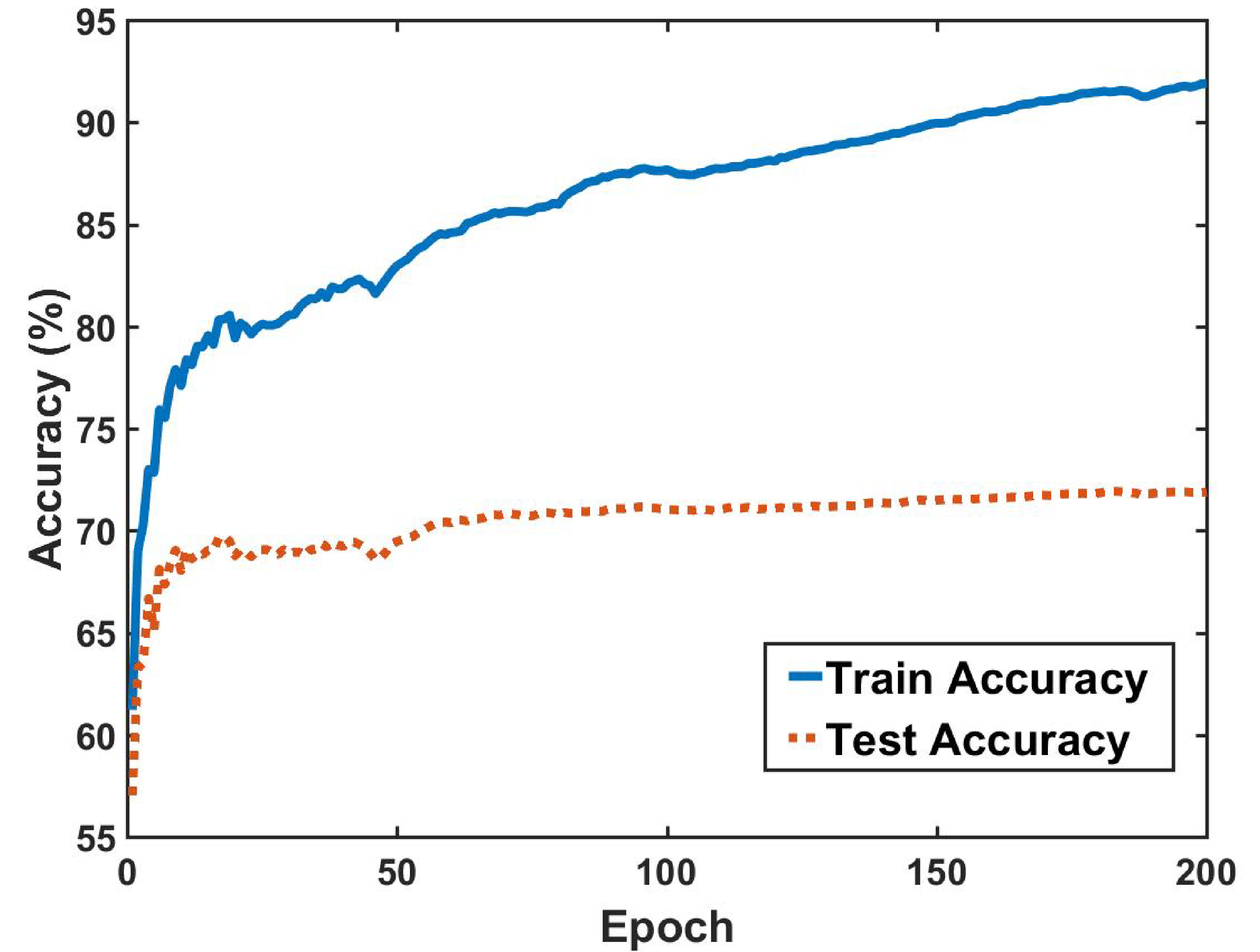}}
\caption{Training accuracy and test accuracy are plotted as a function of the epoch number during training of designed FCNN on MNIST dataset. Accuracy is determined by the number of times the FCNN generates the desired output for the given input.}
\end{figure}

\begin{figure}
\centerline{\includegraphics[width=3.5in]{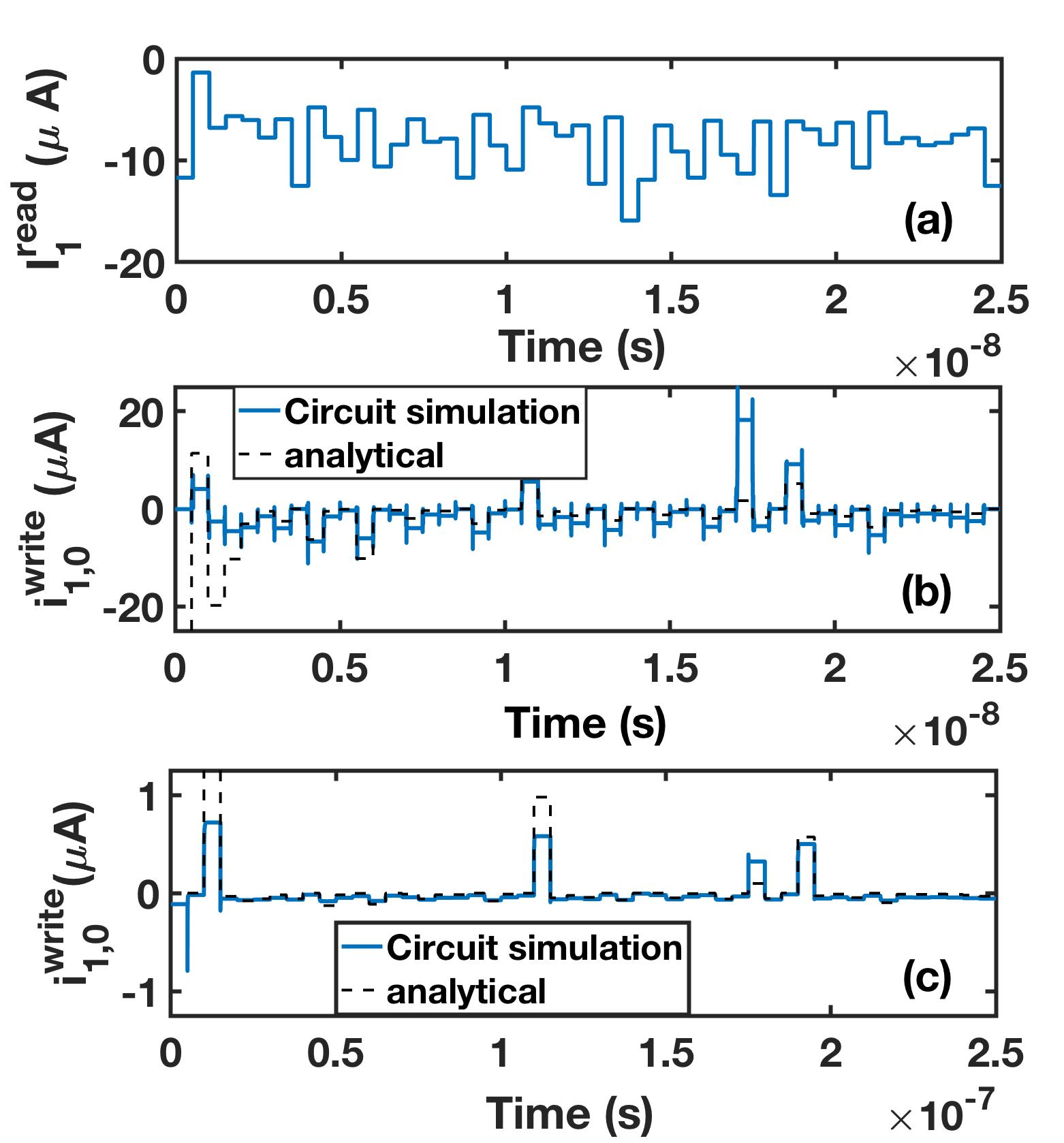}}
\caption{(a)Read current $I^{read}_{1}$ at output node 1, corresponding to digit '0', for the first 50 training samples of first epoch is plotted as function of time. Training for each sample lasts 0.5 ns. (b) Write current  $i^{write}_{1,0}$ generated by circuit simulation of neuron circuit and SGD calculation circuit at node 1 to be sent back to bias synapse at that node is plotted (solid line). Neuron circuit considered is here is combination of op-amp voltage amplifier and transistor differential circuit from Fig.3(a). Training for each sample lasts 0.5 ns. Same   $i^{write}_{1,0}$ obtained analytically  is also plotted (dashed line). (c) Write current  $i^{write}_{1,0}$ generated by circuit simulation of neuron circuit and SGD calculation circuit at node 1 to be sent back to bias synapse at that node is plotted here (solid line), where neuron circuit is differential amplifier based that takes current as input from Fig. 4(a). Training for each sample lasts 5 ns. Thus write current for first 50 training samples of first epoch is shown here as well. Same $i^{write}_{1,0}$ obtained analytically  is also plotted (dashed line).}
\end{figure}

\begin{figure}
\centerline{\includegraphics[width=2.5in]{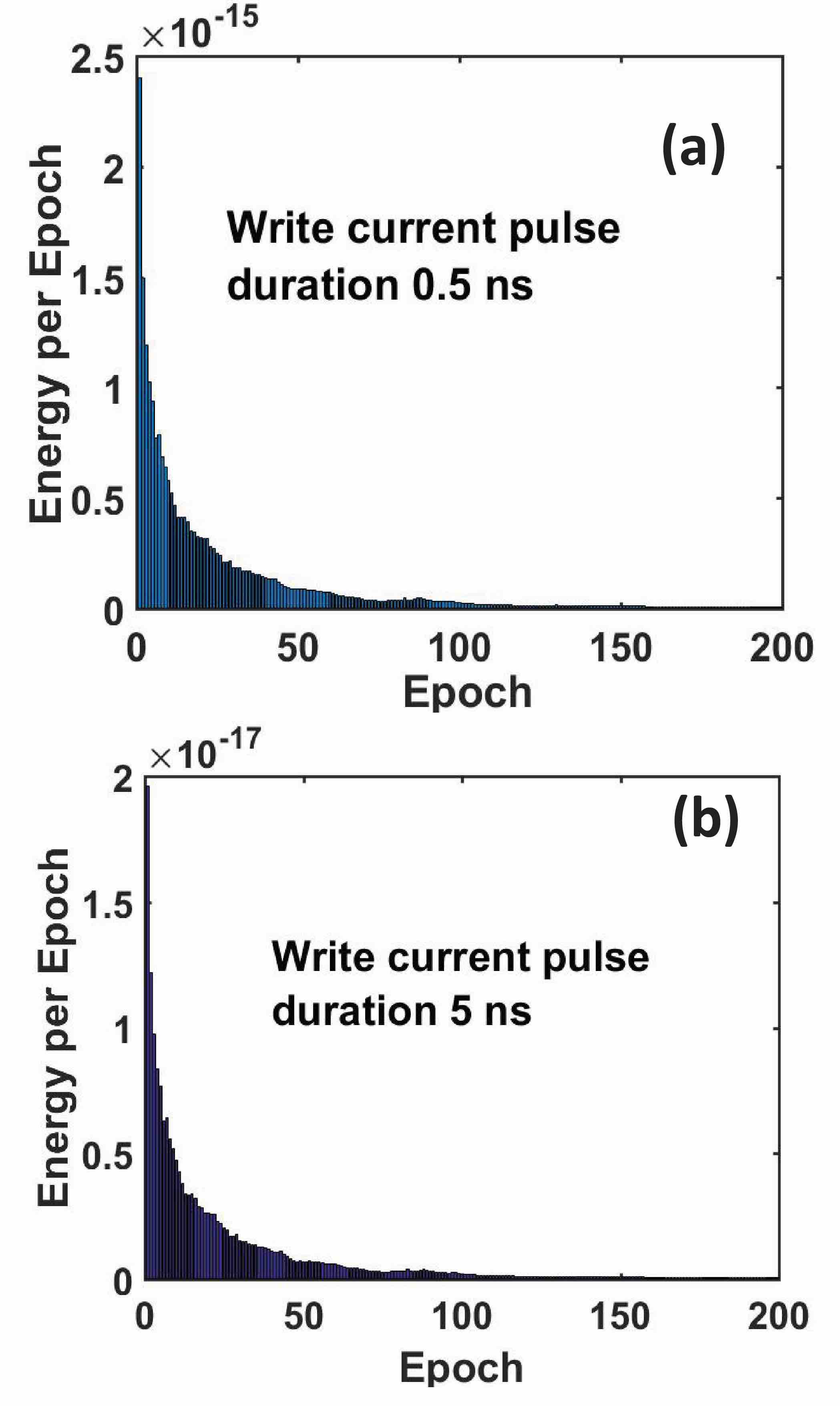}}
\caption{(a) Energy dissipation across all synaptic devices per epoch as a function of epoch number during on-chip learning of designed FCNN on MNIST dataset, when duration of write current pulse is 0.5 ns. (b) Energy dissipation across all synaptic devices per epoch as a function of epoch number when duration of of write current pulse is 5 ns.}
\end{figure}

In order to train the FCNN here on MNIST dataset in software, equations (2), (10), (11) and (12) are solved iteratively over several epochs in numerical package- Python. For 5000 examples in the training set and 10,000 examples in the test set, accuracy is plotted as a function of number of epochs in Fig. 6. After 200 epochs, the training accuracy is $\approx$ 92 percent. Thus the network has been very well trained on the training set. Testing accuracy turns out to be $\approx$ 72 percent, which can be further improved by inserting hidden layers in the network \cite{Haykin,HiddenLayerTheory1,HiddenLayerTheory2}. However, if we insert hidden layers in our FCNN, the weight update for next iteration  expression of equation (11) and (12) will depend upon the weights of the synapses at the different layers for the present iteration after applying chain rule in equations (11) and (12) \cite{LeCun}. In hardware since weight of the synapse is stored as its conductance, its value can only be retrieved by passing a current through it which the feedforward circuit can do but the feedback circuit for weight update cannot (Fig. 1(b)). Hence we do not insert hidden layers in our FCNN.

The maximum magnitude that weight of any synapse in the network takes during the training process is also obtained and used in the corresponding equations for hardware as $w_{max}$. The equations for hardware training of the FCNN (equations  3 - 14) are next solved iteratively in Python. Read currents at the output layer $I^{read}$ and write currents sent by the feedback circuit to the synapses $i_{write}$ are obtained as a function of iteration. Considering that this training happens real time in hardware and time duration of every iteration is equal to the duration of write current pulse for the synaptic device $I^{read}$ and $i_{write}$ are next obtained as function of time. We call this our analytical result. Now, the neuron circuits of Fig. 3(a) and Fig. 4(a) are expected to execute equations (8) and (9) and the SGD calculating feedback circuit of Fig. 5 is expected to execute equation (10)-(14). In order to make sure these circuits we design work as expected, $I^{read,n}$ obtained analytically for the case of training of MNIST at output node $n$ is fed to the neuron circuit of Fig. 3(a) followed by SGD calculation circuit of Fig. 5. Duration of write current pulse to synaptic devices is taken to be 0.5 ns (Fig. 2(b)). Hence every iteration lasts 0.5 ns. Time dependent simulation of the circuits is carried out on Cadence Virtuoso simulator. Write current generated by the SGD circuit $i^{write}_{n,m}$ that will be fed to synapse connecting input node $m$ with output node $n$ is obtained from the circuit simulation and compared with analytical result. 

Fig. 7(a) shows such read current waveform at node 1, corresponding to digit '0' - $I^{read}_{1}$ for a certain time window: 0 to 25 ns, which essentially corresponds to first 50 iterations, i.e. first 50 training samples in the first epoch. The corresponding write current at bias synapse connected to node 0 ( $i^{write}_{1,0}$ ) obtained analytically as well as through circuit simulations is plotted in Fig. 7(b). We observe significant match between analytical and circuit simulation results. The same process is repeated to obtain results in Fig. 7(c), just that the circuit of Fig. 4(a) is used as neuron instead of the circuit of Fig. 3(a) and duration of each iteration is taken to be 5 ns, corresponding to a 5 ns long "write" current pulse. As a result smaller magnitude of write current is needed to bring about the required weight update.  We observe significant match between analytical and circuit simulation results in this case too, showing that we indeed have been able to design the complete network in hardware to carry out "on-chip" learning on the MNIST dataset.

From $i_{write}$ obtained analytically as a function of iterations during the training process for a synaptic device connecting output node $n$ with input node $m$, the corresponding heat energy ($E^{write}$) dissipated in the heavy metal layer of the domain wall synaptic device can be obtained from the following expression:

\begin{equation}
E^{write}_{n,m}=(i^{write}_{n,m})^{2}R_{write}t_{pulse}
\end{equation}

where $t_{pulse} $ is duration of each write current pulse. Considering the heavy metal to be Pt and the device dimensions used in simulations, $R_{write}$ turns out to be 100 $\Omega$. Adding the energies for all the synaptic devices, the total energy dissipation in synaptic devices during weight update is calculated and plotted as function of epochs in Fig. 8. Fig. 8(a) corresponds to pulse of 0.5 ns duration, while Fig. 8(b) corresponds to pulse of 5 ns duration. Since $i_{write}$ in the latter case is lower (Fig. 7) write energy is orders of magnitude lower when pulse duration is longer. Also maximum energy is dissipated during the initial epochs. Once the network starts getting trained, i.e., the accuracies start saturating (Fig. 6) energy dissipation per epoch also reduces because the weights start converging to trained values. Summing over 200 epochs, total energy dissipated in all synaptic devices  is $2.33 \times 10^{-14}$  J for 0.5 ns long pulse and $1.9 \times 10^{-16} $ J for 5 ns long pulse. Since there are 7850 synapses in the network, energy dissipated per synapse for the entire training is as low as $3 \times 10^{-18}$ J for 0.5 ns long pulse and $2.4 \times 10^{-20}$ J for 5 ns long pulse.

\section{Conclusion}

In conclusion, we have designed a feedforward FCNN using domain wall based devices as synaptic devices and transistor based differential amplifier circuits as neurons. We have also designed a feedback circuit through analog electronics that sends write currents to the synaptic devices and updates the corresponding weights. We have simulated the feedforward and feedback circuits together to updates weights of the synapses at every iteration and train the network over the MNIST dataset. We have also reported the performance connected to this "on-chip learning" through training and test accuracy numbers. Hardware limitation connecting to inserting hidden layers limits the test accuracy which can be subject of research for the future. The circuits we design here and the accuracy numbers we report along with the hardware limitations are not only applicable to domain wall synapse based FCNN but also other analog implementations of FCNN, which use other kind of spintronic devices, e.g., skyrmionic devices \cite{Saxena,Skyrmion_WangKang1,Skyrmion_WangKang2}, or non spintronic devices, e.g., 
memristor \cite{memristor} and phase change memory \cite{PCM}, as synapses, if we make slight device specific modifications.

% if have a single appendix:_
% or



\section*{Acknowledgment}

The authors would like to thank Atul Thakur and Prof. A. P. Pratosh, all from Indian Institute of Technology Delhi, for help with simulations and related discussions. This work is partly supported by DST INSPIRE Faculty Fellowship, awarded to Debanjan Bhowmik.

\ifCLASSOPTIONcaptionsoff
  \newpage
\fi

\end{document}